\documentstyle[twocolumn,eqsecnum,aps]{revtex}
\begin{document}
\draft
\title{A 3+1 Dimensional Light-Front Model with
Spontaneous Breaking of Chiral Symmetry}
\author{M. Burkardt and H. El-Khozondar}
\address{Department of Physics\\
New Mexico State University\\
Las Cruces, NM 88003-0001\\U.S.A.}
\maketitle
\begin{abstract}
We investigate a 3+1 dimensional toy model that exhibits spontaneous
breakdown of chiral symmetry, both in a light-front (LF)
Hamiltonian and in a Euclidean Schwinger-Dyson (SD) formulation.
We show that both formulations are completely equivalent ---
provided the renormalization is properly done. For the model considered,
this means that if one uses the same transverse momentum cutoff
on the SD and LF formulations then
the vertex mass in the LF calculation must be taken to be the same 
as the current
quark mass in the SD calculation. The kinetic mass term in the LF
calculation is renormalized non-trivially, which is eventually responsible
for the mass generation of the physical fermion of the model.
\end{abstract}
\narrowtext
\section{Introduction}
Light-front (LF) coordinates are natural coordinates for describing
scattering processes that involve large momentum transfers ---
particularly deep inelastic scattering. This is because correlation 
functions at high energies are often dominated by the free quark
singularities which are along light-like direction.
This is one of the main motivations for formulating field theories
in the LF framework \cite{mb:adv}. But light-front field theories have
another peculiar feature, namely naive reasoning suggests that
the vacuum of all LF Hamiltonians is equal to the Fock vacuum
\cite{ho:vac,mb:adv}. It thus {\it appears} as if LF Hamiltonians
(at least those without the so-called zero-modes, i.e. modes
with $k^+=0$)
cannot be able to describe theories where the vacuum has any nontrivial
features, such as QCD --- where chiral symmetry is believed to
be spontaneously broken. Even if one is only interested in parton 
distributions, one might be worried about using a framework where
the vacuum is just empty space to describe a theory like QCD.

However, it is not quite so easy to dismiss LF field theory as
the following few examples show: One of the first field theories that
was completely solved in the LF formulation was $QCD_{1+1}(N_C\rightarrow
\infty)$ \cite{thooft}. 't Hooft's solution did not include any
zero-modes and therefore he had a trivial vacuum. 
Nevertheless, his spectrum agreed perfectly well with the numerical
results from calculations based on equal time Hamiltonians \cite{wilets}.
Beyond that, application of current algebra sum rules to spectrum
and decay constants obtained from the LF calculation formally gave
nonzero values for the quark condensates that also agreed with numerical results at equal time \cite{zhit}. 
This peculiar result could be understood by defining
LF field theory through a limiting procedure, which showed that some
observables (here: spectrum and decay constants) have a smooth and continuous 
LF limit, while others (here quark condensates) have a discontinuous LF limit.
Other examples have been  studied, in which it was still possible
to demonstrate equivalence between LF results and equal time results 
nonperturbatively, provided the LF Hamiltonian was appropriately renormalized
\cite{mbsg,fr:eps,mb:parity}. Even though these examples are just 1+1
dimensional field theories, it is generally believed among the optimists in 
the field \cite{all:lftd,dgr:elfe,mb:adv} that it should be possible
in 3+1 dimensional field theories as well to achieve equivalence between
spectra of LF Hamiltonian and equal time Hamiltonians by appropriate
renormalization. However, no nontrivial examples (examples that
go beyond mean field calculations) to support such a belief existed
so far.

In this paper, we will give a 3+1 dimensional toy model that can
be solved both in a conventional framework (here by solving the
Schwinger-Dyson equations) as well as in the LF framework.
We will unashamedly omit zero modes as explicit degrees of freedom
throughout the calculation. Nevertheless, we are able to show
that appropriate counter-terms to the LF Hamiltonian 
are sufficient to demonstrate equivalence of the spectrum and other 
physical properties between the two frameworks.

\section{A Simple Toy Model}
The model that we are going to investigate consists of fermions
with some ``color'' degrees of freedom (fundamental representation)
coupled to the transverse component of a vector field, which also
carries ``color'' (adjoint representation). The vector field does not
self-interact.\footnote{Note that, for finite $N_C$, box diagrams with
fermions would induce four boson counter-terms, which we will ignore
here since we will consider the model only in the large $N_C$ limit.}
Furthermore, we will focus on the limit of an infinite number
of colors: $N_C\rightarrow \infty$ ($g$ fixed), which will render the model solvable in the Schwinger-Dyson approach
\begin{equation}
{\cal L} = \bar{\psi}\left( i \partial\!\!\!\!\!\!\not \;\; - m -
\frac{g}{\sqrt{N_C}}{\vec \gamma}_\perp {\vec A}_\perp \right)\psi - \frac{1}{2}
\mbox{tr} \!\!\left( {\vec A}_\perp \Box {\vec A}_\perp + 
\lambda^2 {\vec A}_\perp^2\right).
\end{equation}
With ``$\perp$ component'' we mean here the $x$ and $y$ components.
Furthermore we will impose a transverse momentum cutoff on the fields
and we will consider the model at fixed cutoff.
Also, even though we are interested in the chiral limit of this
model, we will keep a finite quark mass since the LF formulation
has notorious difficulties in the strict $m=0$ case.
Those difficulties can be avoided if one takes $m>0$ 
considers $m\rightarrow 0$.

Even though certain elements of the model bear some resemblance
to terms that appear in the QCD Lagrangian, the model seems is
a rather bizarre construction. However, there is a reason for
this: What we are interested in is a LF-investigating of a model 
that exhibits spontaneous breakdown of chiral symmetry. Furthermore,
we wanted to be able to perform a ``reference calculation'' in a
conventional (non-LF) framework. In the large $N_C$ limit, the rainbow
approximation for the fermion self-energy becomes exact, which allows
us to solve the model exactly in the Schwinger-Dyson
approach. The vector coupling of the bosons to the fermions
was chosen because it is chirally invariant and because a similar
coupling occurs in regular QCD. The restriction to the $\perp$
component of the fields avoids interactions involving
``bad currents''.
Finally, using a transverse momentum cutoff both in the 
Schwinger-Dyson approach and in the LF calculation should allow us
to directly compare the two formulations.

\subsection{Schwinger-Dyson Solution}
Because the above toy model lacks full covariance 
(there is no symmetry relating longitudinal and transverse coordinates)
the full fermion propagator is of the form
\begin{equation}
S_F(p^\mu) = \not \! \! p_L S_L({\vec p}_L^2,{\vec p}_\perp^2)
+ \not \! \! \;p_\perp S_\perp({\vec p}_L^2,{\vec p}_\perp^2)+S_0({\vec p}_L^2,{\vec p}_\perp^2),
\end{equation}
where $\not \! \! k_L \equiv k_0\gamma^0 + k_3 \gamma^3$ and
$\not \! \! k_\perp \equiv k_1\gamma^1 + k_2 \gamma^2$. 
On very general grounds, it should always be possible to write
down a spectral representation for $S_F$\footnote{What we need
is that the Green's functions are analytic except for poles
and that the location of the poles are consistent with
longitudinal boost invariance (which is manifest in our model).
The fact that the model is not invariant under transformations
which mix $p_L$ and $p_\perp$ does not prevent us from writing
down a spectral representation for the dependence on $p_L$.
}
\begin{equation}
S_i({\vec p}_L^2,{\vec p}_\perp^2) = \int_0^\infty dM^2 \frac{\rho_i(M^2,{\vec p}_\perp^2)}
{{\vec p}_L^2-M^2+i\varepsilon},
\label{eq:sansatz}
\end{equation}
where $i=L,\perp,0$.
Note that this spectral representation differs from what one
usually writes down as a spectral representation in that we are not 
assuming full covariance here.
Note that in a covariant theory, one usually writes down spectral
representations in a different form, namely
$S=\int_0^\infty d\tilde{M}^2 \tilde{\rho}(\tilde{M}^2)/({\vec p}_L^2-{\vec p}_\perp^2-\tilde{M}^2)$, i.e.
with ${\vec p}_\perp^2$ in the denominator. This is a special case of
Eq. (\ref{eq:sansatz}) with $\rho(M^2,{\vec p}_\perp^2)=
\int_0^\infty d\tilde{M}^2 \tilde{\rho}
(\tilde{M}^2)\delta(M^2-\tilde{M}^2-{\vec p}_\perp^2)$.

Using thise above ansatz (\ref{eq:sansatz})
for the spectral densities, one finds for the 
self-energy
\begin{eqnarray}
\Sigma(p^\mu) &\equiv& ig^2 \int \frac{d^4k}{(2\pi )^4} {\vec \gamma}_\perp
S_F(p^\mu-k^\mu){\vec \gamma}_\perp \frac{1}{k^2-\lambda^2+i\varepsilon}
\nonumber\\
&=& \not \! \! p_L\Sigma_L({\vec p}_L^2,{\vec p}_\perp^2) + 
\Sigma_0({\vec p}_L^2,{\vec p}_\perp^2),
\label{eq:sd1}
\end{eqnarray}
where
\begin{eqnarray}
\Sigma_L({\vec p}_L^2,{\vec p}_\perp^2) &=& g^2 \!\!\int_0^\infty \!\!\!\!\!dM^2 \!\int_0^1\!\!\!dx 
\!\!\!\int \!\frac{d^2k_\perp}{8\pi^3} 
\frac{(1-x)\rho_L(M^2, ({\vec p}-{\vec k})_\perp^2)}{D}
\nonumber\\
\Sigma_0({\vec p}_L^2,{\vec p}_\perp^2) &=& -g^2 \!\!\int_0^\infty \!\!\!\!\!dM^2 \!\int_0^1\!\!\!dx 
\!\int \!\frac{d^2k_\perp}{8\pi^3} 
\frac{\rho_0(M^2, ({\vec p}-{\vec k})_\perp^2)}{D}.
\nonumber\\
\label{eq:sd2}
\end{eqnarray}
and
\begin{equation}
D=x(1-x){\vec p}_L^2 - xM^2
-(1-x)\left({\vec k}_\perp^2+\lambda^2\right)
\end{equation}
Note that $\Sigma_\perp$ vanishes, since $\sum_{i=1,2} \gamma_i \gamma_j
\gamma_i=0$ for $j=1,2$.
Self-consistency then requires that
\begin{equation}
S_F = \frac{1}{\not \! \! p_L\left[1-\Sigma_L({\vec p}_L^2,{\vec p}_\perp^2) \right]
+ \not \! \! p_\perp - \left[m+\Sigma_0({\vec p}_L^2,{\vec p}_\perp^2)\right]}
\label{eq:sd3}
\end{equation}
In the above equations we have been sloppy about cutoffs in order
to keep the equations simple, but this can be easily remedied by
multiplying each integral by a cutoff on the fermion momentum, such as
$\Theta\left(\Lambda^2_\perp-({\vec p}-{\vec k})_\perp^2\right)$
In principle, the set of equations 
[Eqs. (\ref{eq:sd1}),(\ref{eq:sd2}),(\ref{eq:sd3})]
can now be used to determine the spectrum of the model.
But we are not going to do this here since we are more interested
in the LF solution to the model. However, we would still like
to point out that, for large enough $g$, one obtains a self-consistent
numerical solution to the Euclidean version of the model which
has a non-vanishing scalar piece --- even for vanishing current
quark mass $m$, i.e. chiral symmetry is spontaneously
broken and a dynamical mass is generated for the fermion in this model.
\subsection{LF Solution}
A typical framework that people use when solving LF quantized
field theories is discrete light-cone quantization (DLCQ) 
\cite{pa:dlcq}. Since it is hard to take full advantage of the
large $N_C$ limit in DLCQ, we prefer to use a Green's function 
framework based on a 4 component formulation of the model. 

In a LF formulation of the model, the fermion propagator
(to distinguish the notation from the one above, we denote
the fermion propagator by $G$ here) should be of the form
\footnote{Note that in a LF formulation, $G_+$ and $G_-$ are
not necessarily the same.}
\begin{eqnarray}
G(p^\mu) &=& \gamma^+ p^-G_+(2p^+p^-,{\vec p}_\perp^2)
+\gamma^- p^+G_-(2p^+p^-,{\vec p}_\perp^2)
\nonumber\\
& &+ \not \!\!k_\perp G_\perp(2p^+p^-,{\vec p}_\perp^2)+
G_0(2p^+p^-,{\vec p}_\perp^2).
\label{eq:lf1}
\end{eqnarray}
Again we can write down spectral representations
\begin{eqnarray}
G_i(2p^+p^-,{\vec p}_\perp^2) = \int_0^\infty dM^2
\frac{\rho_i^{LF}(M^2,{\vec p}_\perp^2)}
{2p^+p^--M^2+i\varepsilon},
\label{eq:speclf}
\end{eqnarray}
where $i=+,-,\perp,0$. This requires some explanation:
On the LF, one might be tempted to allow for two terms
in the spectral decomposition of the term proportional to
$\gamma^+$, namely
\begin{equation}
tr(\gamma^-G)\propto
\int_0^\infty dM^2
\frac{p^-\rho_a(M^2,{\vec p}_\perp^2)+\frac{1}{p^+}\rho_b(M^2,{\vec p}_\perp^2)}
{2p^+p^--M^2+i\varepsilon}.
\label{eq:rhoab}
\end{equation}
However, upon writing 
\begin{equation}
\frac{1}{p^+}=\frac{1}{p^+M^2}\left(M^2-2p^+p^-\right)+\frac{2p^-}{M^2}
\end{equation}
one can cast Eq. (\ref{eq:rhoab}) into the form
\begin{eqnarray}
tr(\gamma^-G)&\propto&
\int_0^\infty dM^2p^-
\frac{\rho_a(M^2,{\vec p}_\perp^2)+\frac{2}{M^2}\rho_b(M^2,{\vec p}_\perp^2)
}{2p^+p^--M^2+i\varepsilon}
\nonumber\\
& &-\frac{1}{p^+}\int_0^\infty dM^2
\frac{\rho_b(M^2,{\vec p}_\perp^2)}{M^2},
\label{eq:rhoaab}
\end{eqnarray}
which is of the form in Eq.(\ref{eq:speclf}) plus an energy independent
term. The presence of such an additional energy independent
term would spoil the high energy behavior of the model \cite{brazil}:
In a LF Hamiltonian, not all coupling constants are arbitrary.
In many examples, 3-point couplings and the 4-point couplings
must be related to one another so that the high energy behavior
of scattering via the 4-point interaction and via the iterated
3-point interaction cancel \cite{brazil}. If one does not
guarantee such a cancellation then the high-energy behavior of the
LF formulation differs from the high-energy behavior in covariant
field theory and in addition one often also gets a spectrum that is unbounded
from below. In Eq. (\ref{eq:rhoaab}), the energy independent
constant appears if the coupling constants of the "instantaneous 
fermion exchange" interaction in the LF Hamiltonian and the 
boson-fermion vertex are not properly balanced.
In the following we will assume that one has started with an
ansatz for the LF Hamiltonian with the proper high-energy behavior,
i.e. we will assume that there is no such energy independent 
piece in Eq. (\ref{eq:rhoaab}).

The LF analog of the self-energy equation is obtained by
starting from an expression similar to Eq.(\ref{eq:sd2}) and
integrating over $k^-$. One obtains
\begin{equation}
\Sigma^{LF} = \gamma^+\Sigma_+^{LF}+\gamma^-\Sigma_-^{LF}
+\Sigma_0^{LF},
\end{equation}
where
\begin{eqnarray}
\!\Sigma_+^{LF}\!(p) \!&=&\! g^2 \!\!\!\int_0^\infty \!\!\!\!\!\!dM^2 
\!\!\!\int_0^{p^+}\!\!\!\!\!\!\!dk^+ 
\!\!\!\!\int \!\!\frac{d^2k_\perp}{16\pi^3} 
\frac{\!\!\left(\!p^-\!\!-\frac{\lambda^2+{\vec k}_\perp^2}
{2k^+}\!\right)\!\rho_+^{LF}(M^2\!\!, 
({\vec p}-{\vec k})_\perp^2)}{k^+(p^+-k^+)D^{LF}}
\nonumber\\
& &+ CT
\nonumber\\
\!\Sigma_-^{LF}\!(p) \!&=&\! g^2 \!\!\!\int_0^\infty \!\!\!\!\!\!dM^2 \!\!\!\int_0^{p^+}\!\!\!\!\!\!\!dk^+
\!\!\!\!\int \!\!\frac{d^2k_\perp}{16\pi^3} 
\frac{\left(p^+-k^+\right)\rho_-^{LF}(M^2, ({\vec p}-{\vec k})_\perp^2)}{k^+(p^+-k^+)D^{LF}}
\nonumber\\
\!\Sigma_0^{LF}\!(p) \!&=&\! -g^2 \!\!\!\int_0^\infty \!\!\!\!\!\!dM^2 \!\!\!\int_0^{p^+}\!\!\!\!\!\!\!dk^+
\!\!\!\!\int \!\!\frac{d^2k_\perp}{16\pi^3} 
\frac{\rho_0^{LF}(M^2, ({\vec p}-{\vec k})_\perp^2)}{k^+(p^+-k^+)D^{LF}}.
\label{eq:lf2}
\end{eqnarray}
where
\begin{equation}
D^{LF}=p^- - \frac{M^2}{2(p^+-k^+)} - \frac{\lambda^2+{\vec k}_\perp^2}{2k^+}
\end{equation}
and CT is an energy ($p^-$)
independent counter-term. The determination of this counter-term, such
that one obtains a complete equivalence with the Schwinger Dyson 
approach, is in fact the main achievement of this paper.
First we want to make sure that the counter-term renders the self-energy
finite. This can be achieved by performing a ``zero-energy subtraction''
with a free propagator,
analogous to adding self-induced inertias to a LF Hamiltonian, yielding
\begin{equation}
CT= g^2 \int_0^{p^+}\!\!\!\!\!\!\!dk^+ 
\!\!\!\!\int \!\!\frac{d^2k_\perp}{16\pi^3} 
\frac{\!\!\frac{\lambda^2+{\vec k}_\perp^2}{2k^+}\!}{k^+(p^+-k^+)D_0^{LF}}
+\frac{\Delta m^2_{ZM}}{2p^+},
\label{eq:ct1}
\end{equation}
where
\begin{equation}
D_0^{LF}= - \frac{M_0^2+({\vec p}-{\vec k})_\perp^2}{2(p^+-k^+)} - \frac{\lambda^2+{\vec k}_\perp^2}{2k^+}
\end{equation}
and where we denoted the finite piece by $\Delta m^2_{ZM}$ (for {\it zero-mode}), since
we suspect that it arises from the dynamics of the zero-modes.
$M_0^2$ is an arbitrary scale parameter. We will construct
the finite piece ($\Delta m^2_{ZM}$) so that there is no 
dependence on $M_0^2$ lect in CT in the end.

At this point, only the infinite part of $CT$ is unique \cite{brazil}, since it
is needed to cancel the infinity in the $k^+$ integral
in Eq. (\ref{eq:lf2}), while the
finite (w.r.t. the $k^+$ integral) piece (i.e. $\Delta m^2_{ZM}$) seems arbitrary. \footnote{Note that what we called the "finite piece"
w.r.t. the $k^+$ integral is still divergent when one integrates over 
$d^2k_\perp$ without a cutoff!}
Below we will show that it is not arbitrary and only
a specific choice for $\Delta m^2_{ZM}$
leads to agreement between the SD and the LF approach.

Note that the equation for the self-energy can also be written in the
form
\begin{eqnarray}
\!\Sigma_+^{LF}\!(p) \!&=&\! g^2  
\!\!\!\int_0^{p^+}\!\!\!\!\frac{dk^+}{k^+} 
\!\!\!\!\int \!\!\frac{d^2k_\perp}{8\pi^3} 
p^-_F
G_+\left(2p^+_Fp^-_F,{\vec p}_{\perp F}^2\right)
+ CT
\nonumber\\
\!\Sigma_-^{LF}\!(p) \!&=&\! g^2  
\!\!\!\int_0^{p^+}\!\!\!\!\frac{dk^+}{k^+}
\!\!\!\!\int \!\!\frac{d^2k_\perp}{8\pi^3} 
p^+_F
G_-\left(2p^+_Fp^-_F,{\vec p}_{\perp F}^2\right)
\nonumber\\
\!\Sigma_0^{LF}\!(p) \!&=&\! -g^2 
\!\!\!\int_0^{p^+}\!\!\!\!\frac{dk^+}{k^+}
\!\!\!\!\int \!\!\frac{d^2k_\perp}{8\pi^3} 
G_0\left(2p^+_Fp^-_F,{\vec p}_{\perp F}^2\right),
\label{eq:lf2b}
\end{eqnarray}
where
\begin{eqnarray}
p^+_F&\equiv& p^+-k^+ \nonumber\\
p^-_F&\equiv& p^--\frac{\lambda^2+{\vec k}_\perp^2}{2k^+}\nonumber\\
{\vec p}_{\perp F} &\equiv&{\vec p}_\perp-{\vec k}_\perp
\end{eqnarray}
One can prove this by simply comparing expressions! 
Bypassing the use of the spectral function greatly simplifies
the numerical determination of the Green's function in a self-consistent
procedure.

\subsection{DLCQ solution}
There are reasons why one might be sceptical about the
Green's function approach to the LF formulation of the model:
First we used a four-component formulation which resembles
a covariant calculation. Furthermore, we introduced spectral 
representations for the Green's functions and assumed certain
properties [Eq.(\ref{eq:speclf})].
Since we were initially also sceptical, we performed the following
calculation: First we formulated the above model as a Hamiltonian
DLCQ problem \cite{pa:dlcq} with anti-periodic boundary conditions
for the fermions and periodic boundary conditions for the bosons in the
longitudinal direction. Zero modes ($k^+=0$) were omitted. 
This is a standard procedure and we will not give
any details here. The only nontrivial step was the choice of
the kinetic energy for the fermion, which we took, using
Eq. (\ref{eq:ct1}), to be
\begin{equation}
T=\sum_{{\vec p}_\perp}\sum_{p^+=1,3,..}^{\infty} T(p)\left(b^\dagger_pb_p
+d^\dagger_pd_p\right),
\end{equation}
with
\begin{eqnarray}
T(p)&=&\frac{m^2+{\vec p}_\perp^2+\Delta m^2_{ZM}}{p^+} \\
& &+ \sum_{{\vec q}_\perp} \sum_{q^+=1,3,..}^{p^+-2}
\frac{1}{ {q^+}^2(p^+-q^+)}\frac{m^2+{\vec q}_\perp^2}{\frac{\lambda^2+({\vec p}_\perp-{\vec q}_\perp)^2}{k^+-q^+}
+\frac{m^2+{\vec q}_\perp^2}{q^+}}
\nonumber
\end{eqnarray}
(some cutoff, such as a sharp momentum cutoff, is implicitly assumed).
Having obtained the eigenvalues of the DLCQ Hamiltonian,
we then determined the Green's function self-consistently,
by iteratively solving Eq. (\ref{eq:lf2b}),
using the same cutoffs as in the DLCQ calculation:
the same transverse momentum cutoff and discrete $k^+$ summations instead
of the integrals.
The result was that the invariant mass at the first pole of the
self-consistently determined Green's function coincides to at least
10 significant digits (!) with the invariant mass of the physical fermion
as determined from the DLCQ diagonalization.
This result was independent of the cutoff --- as long as the same
cutoff was used in both the Green's function and the DLCQ approach.
This proves that the self-consistent Green's function calculation and
the DLCQ calculation are in fact completely equivalent for our toy model. This is a very useful result, since it allows us to formally 
perform the continuum limit (replace sums by integrals)
--- a step that is clearly impossible for a DLCQ calculation.

\subsection{Comparing the LF and SD solutions}
Having established the equivalence between the Green's function method and
the DLCQ approach, we can now proceed to compare the Green's function
approach (in the continuum) with the Schwinger-Dyson approach.
Motivated by considerations in Ref.\cite{mb:adv}, we make the
following ansatz for ZM:
\begin{equation}
\Delta m^2_{ZM} = g^2\int_0^\infty \!\!\!\!dM^2 \!\!\int 
\!\!\frac{d^2k_\perp}{8\pi^3}
\rho_+^{LF}(M^2,{\vec p}_{F\perp}^2) \ln \frac{M^2}{M_0^2+{\vec p}_{F\perp}^2}.
\label{eq:zm}
\end{equation}
The motivation for this particular ansatz becomes obvious one
we rewrite the expression for $\Sigma_+^{LF}$:
For this purpose, we first note that
\begin{eqnarray}
\frac{p^--\frac{\lambda^2+{\vec k}_\perp^2}{2k^+}}{k^+(p^+-k^+)D^{LF}}
&+&
\frac{\frac{\lambda^2+{\vec k}_\perp^2}{2k^+}}{k^+(p^+-k^+)D_0^{LF}}
\\
= \frac{p^-\frac{p^+-k^+}{p^+}}{k^+(p^+-k^+)D^{LF}}
&-& \frac{1}{p^+}\frac{\partial}{\partial k^+} \ln \left[\frac{D^{LF}}{D_0^{LF}}\right].
\nonumber
\end{eqnarray}
Together with the normalization condition\\ 
$\int_0^\infty dM^2 
\rho_+^{LF}(M^2,{\vec k}_\perp^2)=1$, this implies
\begin{eqnarray}
\!\Sigma_+^{LF}\!(p) \!&=&\! g^2\frac{p^-}{p^+} \!\!\!\int_0^\infty \!\!\!\!\!\!dM^2 
\!\!\!\int_0^{p^+}\!\!\!\!\!\!\!dk^+ 
\!\!\!\!\int \!\!\frac{d^2k_\perp}{16\pi^3} 
\frac{\!\!\left(p^+-k^+\right)\!\rho_+^{LF}(M^2\!\!, ({\vec p}-{\vec k})_\perp^2)}{k^+(p^+-k^+)D^{LF}}
\nonumber\\
& &-\frac{g^2}{2p^+} \int_0^\infty \!\!\!\!\!\!dM^2 \!\!\!\int \!\!\frac{d^2k_\perp}{8\pi^3} 
\rho_+^{LF}(M^2,{\vec p}_{F\perp}^2)\ln \frac{M^2}{M_0^2+{\vec p}_{\perp F}^2 } \nonumber\\
& &+\frac{\Delta m^2_{ZM}}{2p^+}
\nonumber\\
&=&\! g^2\frac{p^-}{p^+} \!\!\!\int_0^\infty \!\!\!\!\!\!dM^2 
\!\!\!\int_0^{p^+}\!\!\!\!\!\!\!dk^+ 
\!\!\!\!\int \!\!\frac{d^2k_\perp}{16\pi^3} 
\frac{\!\!\left(p^+-k^+\right)\!\rho_+^{LF}(M^2\!\!, ({\vec p}-{\vec k})_\perp^2)}{k^+(p^+-k^+)D^{LF}},\nonumber\\
\end{eqnarray}
where we used our particular ansatz for $\Delta m^2_{ZM}$ [Eq. (\ref{eq:zm})].
Thus, for our particular choice for the finite piece of the kinetic
energy counter term, the expression for $\Sigma_+^{LF}$ and $\Sigma_-^{LF}$
are almost the same --- the only difference being the replacement of
$\rho_+^{LF}$ with $\rho_-^{LF}$ and an overall factor of $p^-/p^+$.
Furthermore, the most important result of this paper is
a direct comparison (take $x=k^+/p^+$) shows that the same spectral
densities that provide a self-consistent solution to the SD
equations (\ref{eq:sd2}) also yield a self-consistent solution to the
LF equations, provided one chooses
\begin{eqnarray}
\rho_+^{LF}(M^2,{\vec k}_\perp^2) &=&\rho_-^{LF}(M^2,{\vec k}_\perp^2)
=\rho_L(M^2,{\vec k}_\perp^2)\nonumber\\
\rho_0^{LF}(M^2,{\vec k}_\perp^2) &=&\rho_0(M^2,{\vec k}_\perp^2).
\end{eqnarray}
In particular, the physical masses
of all states (in the sector with fermion number one)
must be the same in the SD and the LF framework.

In the formal considerations above, we found it convenient to
express $\Delta m^2_{ZM}$ in terms of the spectral density.
However, this is not really necessary since one can express
it directly in terms of the Green's function
\begin{eqnarray}
\Delta m^2_{ZM}&=&g^2p^+\!\!\!\!\int_{-\infty}^0\!\!\!\!\!\!dp^-\!\!\!\!
\left.\int \!\!\frac{d^2p_\perp}{
4\pi^3} \!\!\right[
G_+(2p^+p^-,{\vec p_\perp}^2)
\label{eq:dmgreen}
\\
& &\quad \quad \quad \quad \quad \quad \quad \quad - \left.\frac{1}
{2p^+p^--{\vec p}_\perp^2-M_0^2}\right] .
\nonumber
\end{eqnarray} 
Analogously, one can also perform a "zero-energy subtraction" in
Eq. (\ref{eq:lf2b}) with the full Green's function, i.e.
by choosing
\begin{equation}
CT=-g^2  
\!\!\!\int_0^{p^+}\!\!\!\!\frac{dk^+}{k^+} 
\!\!\!\!\int \!\!\frac{d^2k_\perp}{8\pi^3} 
\tilde{p}^-_F
G_+\left(2p^+_F\tilde{p}^-_F,{\vec p}_{\perp F}^2\right),
\label{eq:ctilde}
\end{equation}
with $\tilde{p}^-_F=-(\lambda^2+{\vec k}_\perp^2)/2k^+$.
This expression turns out to be very useful when constructing the
self-consistent Green's function solution. 
We used both ans\"atze [Eqs. (\ref{eq:dmgreen}) and 
(\ref{eq:ctilde})] to determine the physical masses of the
dressed fermion. In both cases, numerical agreement with
the solution to the Euclidean SD equations was obtained.

Note that, in a canonical
LF calculation (e.g. using DLCQ) one should avoid expressions
involving $G_+$, since it is the propagator for the unphysical ("bad")
component of the fermion field that gets eliminated by solving
the constraint equation.
However, since the model that we considered has an underlying Lagrangian
which is parity invariant, one can use $G_+=G_-$ for the self-consistent
solution and still use Eq. (\ref{eq:dmgreen}) or
Eq. (\ref{eq:ctilde}) but with $G_+$ replaced by $G_-$.

\section{Summary}
We studied a simple 3+1 dimensional model with ``QCD-inspired''
degrees of freedom which exhibits spontaneous breakdown of chiral
symmetry. The methods that we used were the Schwinger-Dyson
approach, a LF Green's function approach and DLCQ.
The LF Green's function approach was used to ``bridge'' between
the SD and DLCQ formulations in the following sense:
On the one hand, we showed analytically that the LF Green's function
solution to the model is equivalent to the SD approach.
On the other hand we verified numerically that by discretizing the
momentum integrals, that appeared in the LF Green's function approach,
agreement between the LF Green's function approach and DLCQ.
Hence we have shown that the SD solution and the DLCQ solution are equivalent.
This remarkable result implies that even though the LF calculation was done 
without explicit zero-mode 
degrees of freedom, its solution contain the
same physics as the solution to the SD equation --- including dynamical
mass generation for the fermions.

However, we have also shown that the equivalence between the LF approaches and the
SD approach only happens with a very particular choice for the fermion kinetic
mass counter-term in the light-front framework.
Our calculation also showed that
the current quark mass of the SD calculation is to be identified
with the ``vertex mass'' in the LF calculation --- provided the same
cutoffs are being used in both calculations. This result makes 
sense, considering that both the current quark mass and the LF vertex
mass are the only parameters that break chiral symmetry explicitly.
The mass generation for the fermion in the chiral limit of the 
LF calculation occurs through the kinetic mass counter-term (which
does not break chiral symmetry) \cite{all:lftd}.\footnote{We should add
that the ``kinetic mass counter-term'' did depend on the transverse
momentum of the fermion for most cutoffs other than a transverse momentum
on the fermion.}
Our results contradict Ref. \cite{hari}, where it has been {\it ad hoc}
suggested that the renormalized vertex mass remains finite in the chiral
limit to account for spontaneous breaking of chiral symmetry.

Our work presents an explicit 3+1 dimensional example that there is no
conflict between chiral symmetry breaking and trivial LF vacua
provided the renormalization is properly done.
In our formal considerations, we related the crucial finite piece
of the kinetic mass counter-term to the spectral density.
Several alternative determinations (which might be more suitable for a 
practical calculation) are conceivable:
parity invariance for physical observables \cite{mb:parity},
more input (renormalization conditions) such as fitting the fermion
or ``pion'' mass. 

However, one must be careful with this result in the following sense: although we have provided an explicit example which shows
that, even in a 3+1 dimensionsonal model with
$\chi SB$ for $m\rightarrow 0$, LF Hamiltonians without explicit
zero-modes can give the right physics, we are still far from
understanding whether this is possible in full QCD and how
complicated the effective LF Hamiltonian for full QCD needs to be.
More work is necessary to answer these questions.

As an extension of this work we had planned to study the pion in the
chiral limit of a 1+1
dimensional version of this model using the LF framework.\footnote{Even in 1+1 dimensions, one expects a massless boson in the chiral limit because
of $N_C\rightarrow \infty$. One can show this in the SD formalism
since the solution for the self-energy equation for the fermion
also solves the Bethe-Salpeter equation for the pseudoscalar
bound state with zero mass.} 
We were not able to derive an analog of the Green's function equations
for the pion, so we had to resort to a brute force DLCQ calculation.
Numerical convergence, which was acceptable for 
the fermion, was very poor for the pion in the chiral limit, and
we were thus not able to demonstrate that it emerges naturally as a massless
particle. Nevertheless, we expect that other numerical techniques,
which treat the end point behavior of the LF wavefunctions more carefully
than DLCQ, should yield a massless pion for this model.
\acknowledgements
M.B. would like to acknowledge Michael Frank and Craig Roberts for very
helpful discussions on the Schwinger-Dyson solution to the model.
We thank Dave Robertson for critically reading and commenting on
a preliminary version of this paper.
This work was supported by the D.O.E. under contract DE-FG03-96ER40965
and in part by TJNAF.


\begin{references}
\bibitem{mb:adv} M. Burkardt, Advances\ Nucl.\ Phys.\ {\bf 23}, 1 (1996).
\bibitem{ho:vac} K.~Hornbostel,
Phys. Rev. {\bf D45} (1992) 3781;\\
D.G.~Robertson, Phys.\ Rev.\ D\ {\bf 47}, 2549 (1993).
\bibitem{thooft} G.~'t~Hooft, Nucl.\ Phys. B\ {\bf 75}, 461 (1974).
\bibitem{wilets} M.~Li and L.~Wilets, J.\ Phys. G\ {\bf 13}, 915 (1987).
\bibitem{zhit} A.~Zhitnitsky, Phys.\ Lett.\ {\bf 165 B}, 405 (1985);\\
M.~Burkardt, Phys.\ Rev.\ D\ {\bf 53}, 933 (1996).
\bibitem{eps} F.~Lenz et. al., A..\ Phys. {\bf 208}, 1 (1990).
\bibitem{mbsg} M. Burkardt, Phys.\ Rev.\ {\bf D47}, 4628 (1993).
\bibitem{fr:eps} E. V. Prokhvatilov and V. A. Franke,
Sov. J. Nucl. Phys. {\bf 49} (1989) 688;
E. V. Prokhvatilov, H. W. L. Naus and H.-J. Pirner,
Phys.\ Rev.\ D\ {\bf 51}, 2933 (1995); J.P.~Vary, T.J.~Fields and H.-J.~Pirner, Phys.\ Rev.
 D\ {\bf 53}, 7231 (1996).
\bibitem{mb:parity} M. Burkardt, {\it to appear in Phys. Rev. D},
hep-ph/9601289.
\bibitem{all:lftd}
K. G. Wilson et al., Phys.\ Rev.\ D\ {\bf 49}, 6720 (1994).
\bibitem{dgr:elfe} S.~J.~Brodsky and D.~G.~Robertson,
to be published in the proceedings of ELFE 
Summer School on Confinement Physics, Cambridge, England, 22-28 Jul 1995,
hep-ph/9511374. 
\bibitem{brazil} R. J. Perry, invited lectures presented at
`Hadrons 94', Gramado, Brazil, April 1994, hep-ph/9407056.
\bibitem{pa:dlcq} H.-C.~Pauli and S.~J.~Brodsky, Phys.\ Rev.\ D\ 
{\bf 32}, 1993 (1985); 
{\it ibid} 2001 (1985).
\bibitem{hari} W.M.~Zhang and A.~Harindranath, hep-ph/9606347.
\end{references}
\end{document}